# A Brief History of the Study of High Energy Cosmic Rays using Arrays of Surface Detectors

A A Watson[1]

[1] School of Physics and Astronomy, University of Leeds, Leeds, UK

**Abstract.** A brief history of the development of surface detectors for the study of the high-energy cosmic rays is presented. The paper is based on an invited talk given at UHECR2022 held in L'Aquila, 3 – 7 October 2022. In a complementary talk, P Sokolsky discussed the development of the fluorescence technique for air-shower detection.

## 1 The discovery of extensive air showers

Cosmic-ray physicists have always studied the highest-energy particles that exist in nature, even when, in the earliest days, they thought that the ionisation they were detecting at ground-level originated from primary photons. In this article, I will focus mainly, but not exclusively, on the efforts developed to study cosmic rays with energies above ~0.1 EeV, and I will introduce these in a historical manner. We will find that the techniques now widely-used are, often, simply scaled-up versions of what has gone before: those of us active in the field, should be very grateful to the work of a small number of pioneers.

Above an energy of ~100 TeV, the flux of cosmic rays is so low that direct detection with equipment flown on a balloon or a spacecraft is difficult. Instead, one must make use of the secondary particles that the higher-energy cosmic rays produce, which reach ground level and are spread out so that they can be observed, more or less simultaneously, by arrays of detectors. This phenomenon is known as an *extensive air shower*. The separation of the detectors usually grows as cosmic rays of higher energy are studied, largely for financial reasons, and deployment at mountain altitudes is useful for studies at the lower energies.

B Rossi made the first detection of extensive air showers in 1934 while searching for the East-West effect in Eritrea, [1]. He observed the correlated arrival of particles, in widely-separated Geiger counters, at a rate well above that expected by chance, after taking into account the known resolving time of his coincidence circuitry. Rossi named this phenomenon '*sciami*', '*swarms*' in English. Earlier [2], he had found a rapid increase in the rate of coincidences between a triangular arrangement of Geiger counters, as the thickness of lead above the detectors increased. All three counters could not have been triggered by a single particle travelling in a straight line. The results were so surprising that the paper describing them was at first rejected, and only published after W Heisenberg vouched for Rossi's reliability. These results, now known as the *Rossi transition curves*, led K Schmeiser and W Bothe to conclude that showers would also develop in air. They made the first systematic study of this possibility [3] using Geiger counters separated by up to 50 cm, introducing the term *'luftschauer'* to describe what they observed. At around the same time, W Kolhöster, I Matthes and E Weber [4] made similar investigations with counter separations up to 75 m.

In addition to this systematic work, two other earlier observations pointed to showering in air. Firstly, in 1932, P M S Blackett and G Occhialini [5] noted that some of the cascades seen in a cloud chamber triggered by coincidences between a set of Geiger counters, appeared to originate in the air surrounding their instruments. Secondly, in 1935, E Regener and G Pfotzer [6][1] observed a maximum in the rate of coincidences between three vertically-separated Geiger counters as they were carried upwards by a balloon. The peak intensity, at ~14 km above sea-level, was interpreted by Regener in 1938 [7] as being due to the multiplication of electrons in the atmosphere, which he called *'Schauer'*, his thinking being guided by the theoretical work of H Bhabha and W Heitler [8]. Neither Regener nor Rossi appear to have recognised that the same physical mechanisms lay behind their respective observations. Rossi was unable to follow up his work in part because of his central role in creating a new physics department at the University of Padua and, tragically, because of the anti-Semitic laws of the Fascist regime led to him losing his position[2].

Despite the work of Rossi and the two German groups, credit for the discovery of extensive air showers has usually been given to P Auger and his collaborators

---

[1] This maximum is commonly known as the Pfotzer maximum, though this naming is in fact inappropriate: see P Carlson and A A Watson, Hist Geo Space Sci **5** 1 2014
[2] In 1973, a law was introduced by the Italian Government offering a university chair to those Italians who, like Rossi, had been exiled in this manner. Between 1974 and 1980, Rossi held a professorial position in the University of Palermo.

[9] for what was actually another serendipitous discovery arising from the use of the classical method to measure the resolving time of the new coincidence circuitry developed by R Maze [10][3]. Auger appears to have been unaware of the earlier work of Rossi, or of that of the two German groups, and for a while the commonly used abbreviation for extensive air showers, EAS, was interpreted as '*Extensive Auger Showers*' [11]. On the Jungfraujoch (3463 m), Auger and his team were able to increase the counter separations to over 300 m, and later, using cloud chambers and Geiger counters shielded by lead, to make the discovery that penetrating particles (muons) were associated with the showers. Using the new techniques of quantum electrodynamics, Auger et al. argued [12] that the primary particles causing the largest showers must have energies of ~1 PeV: at this time the highest-energy cosmic rays, as inferred from geomagnetic studies, were only ~10 GeV, so this was an extraordinary extension of Nature's energy scale. Auger thought that the primary particles were electrons.

Auger left France in 1939, spending time in Blackett's laboratory in Manchester, *en route* to Chicago before moving to Canada in 1943. There, his team at Chalk River, that led the British-Canadian-French efforts in support of the Manhattan project, included B Pontecorvo, a talented experimentalist as well as an exceptional theorist. Auger's time in Manchester coincided with the presence there of Rossi who was also *en route* to the United States, where he worked first in Chicago, with A H Compton, and later in Cornell, where K Greisen was his first American graduate student. Both Rossi and Greisen joined the Manhattan project at Los Alamos. The discussions that took place while Auger and Rossi were in Manchester surely enhanced Blackett's long-standing interest in the shower phenomenon [13].

## 2 Activities during World War II

Neither Bothe nor Kolhörster were able to carry on their cosmic-ray work after the outbreak of WWII, although Bothe did attend the International Conference on Cosmic Rays held in Chicago in June 1939[4]. While in Chicago, Auger was able to continue his shower studies and, in collaboration with A Rogozinski and M Schein, flew ten Geiger counters, arranged along a 10 m line, to an altitude of ~15 km (~120 g cm$^{-2}$) in January 1943. Four of the counters were close together at the centre of the array [14]. From the observation that coincidences between these central counters were unaccompanied by hits in the more distant ones, Auger deduced that, at this altitude, the shower consisted of a narrow bunch of particles, an observation now easy to understand from what we know about high-energy interactions.

In the Soviet Union, soon after the Battle of Stalingrad had ended in February 1943, D V Skobeltzyn initiated studies of showers in the Pamirs at an altitude of 3860 m. Results obtained in 1946 [15] extended the work of Auger at a similar altitude and were discussed in the context of the models of cascades then available[5].

A novel way of detecting air showers was discussed in the UK during World War II. On 4 September 1939, just after the outbreak of the War, A C B Lovell and J G Wilson (who was later to lead the Haverah Park project) were visiting a radar station in Yorkshire. Watching over the shoulder of a radar operator, they saw echoes on the screen, which the young woman ignored telling them that these were not from aircraft, but were reflections from the ionosphere. This led Lovell to conceive of using radar to detect the ionisation trail left by shower particles. A paper was written, with Blackett, proposing that the *'radio echo technique'* might be used to detect air showers [16]. A letter from T L Eckersley to Blackett [17] soon followed, in which it was pointed out that the recombination time of the electrons adopted in their calculation was too large by a factor of $10^6$ and consequently that the technique was less promising than claimed. This criticism was forgotten – war-work had precedence - and, soon after the War ended, Blackett encouraged Lovell to look for showers using this technique with surplus war equipment that was readily available. Lovell soon discovered that the trails that he and Wilson had spotted in 1939 were in fact due to meteors. In his writings, Lovell makes clear that had the Eckersley letter been recalled after the War, the Jodrell Bank radio astronomy facilities might never have developed [17].

---

[3] During the third European Cosmic Ray Symposium held in Paris in 1972, I was invited one evening, along with others including Arnold Wolfendale, Cormac O'Ceallaigh, and George Wdowczyk, to the home of Roland Maze. During the evening, Wdowczyk asked Maze to recount the story of the discovery of extensive air showers. Speaking very slowly, in French, he made it clear that his task had been to improve the resolving time of the coincidence circuitry then available. This he did and, to measure the resolving time, he used the random coincidence method. There was nothing in what Maze said to suggest that a search for showers had been planned. Rather it seems very likely that this, like Rossi's, was a serendipitous discovery.

[4] Heisenberg organised two symposia on cosmic rays in Berlin in 1941 and 1942 but neither Bothe nor Kolhörster are listed as contributors (*Cosmic Radiation*, edited by W. Heisenberg, translated by T.H. Johnson, Dover Publications 1946)

[5] It is remarkable, that at this crucial time in Soviet history, research in cosmic rays was supported. During a meeting in Pylos in 2004, G T Zatsepin, a student of Skobeltzyn, commented that Skobeltzyn saw work on cosmic rays as a reconnaissance for the study of high-energy interactions. At the time, the possibility that understanding these interactions might yield a source of energy may have looked appealing to political leaders. A H Compton, when seeking support for his latitude studies in the 1930s, had advanced a similar argument.

# 3 Activities in the immediate post-War period: 1945 - 1965

## 3.1 Results from arrays of Geiger counters used at Moscow State University

It will be apparent from the discussions in sections 1 and 2 that the Geiger counter was the major tool used in the discovery of extensive air showers, and this device remained an important technology well into the 1950s, before being superseded by scintillation counters and water-Cherenkov detectors. A key result from the use of Geiger counters was that obtained by G V Kulikov and G B Khristiansen who, in 1958 [18] reported a steepening of the number spectrum of showers at a size of $\sim 10^6$ particles. At this time, the primary energy corresponding to such a size was known only very approximately and was thought to be ~10 PeV. The steepening, what we now call *'the knee'*, was interpreted by Kulikov and Khristiansen as evidence for an astrophysical feature, specifically a demonstration in favour of an extragalactic origin of cosmic rays above 10 PeV.

## 3.2 Impact of the work of the MIT Group

After finishing work on the Manhattan project, Rossi established a powerful cosmic-ray group at MIT. Perhaps unsurprisingly, one of his interests was the study of extensive air showers. Rossi targeted the determination of the direction and energy of the incoming primaries as being of the highest priority. In his autobiography [19], he states that the program which he initiated *"because of the originality of its conception and the significance of it results, ranks amongst the foremost accomplishment of the MIT group."* Many of the instrumental and analysis techniques developed by his team form the bedrock of what we use today.

Rossi was aware of the invention of the scintillation counter in 1944 by S C Curran who, in support of the Manhattan project, had coupled a photomultiplier to a scintillating crystal to measure radioactivity. Rossi saw the possibility of introducing scintillation counters for the study of air showers. However, larger areas were needed and the first MIT detectors were liquid scintillators made of benzene, with the addition of a small amount of para-terphenol to act as a wavelength shifter. Using three liquid-scintillation counters, each of 600 cm$^2$, P Bassi, G Clark and Rossi [20], in a seminal paper, established that the spread of arrival times of the shower particles was sufficiently small to make determination of shower directions by fast-timing feasible. They also showed that, close to the centre of the shower, the electrons preceded the muons, and that an appreciable fraction of the electrons observed at sea-level arose from nucleons having energies above 20 GeV. This paper remains fascinating reading for its original insights, and as a demonstration of how clear physical reasoning, coupled with observations from a small number of detectors, can lead to profound understandings.

The next step was to lay out an array of 11 scintillators on a triangular grid covering 0.17 km$^2$ at the Agassiz station of Harvard University. A lightning strike led to some of the detectors catching fire and these were then replaced with large-area plastic detectors designed and manufactured by the MIT group [21]. The largest event recorded at the Agassiz array had a primary energy of ~5 EeV.

The work at Agassiz was the forerunner for two important shower projects. One was the construction of arrays in Bolivia, using much of the Agassiz instrumentation, first at a site on the Alto Plano (El Alto) at 4200 m (620 g cm$^{-2}$), and then at an altitude of 5200 m (500 g cm$^{-2}$) on Mt Chacaltaya. At the El Alto site, important measurements of shower development were made using what is now known as the *Constant Intensity Cut method* [22]. The work at El Alto led to the first estimate of the maximum of shower development, ~600 g cm$^{-2}$ at ~0.1 EeV, in reasonable accord with later direct measurements made using the fluorescence technique. In addition to measuring the energy spectrum of cosmic rays to beyond 0.1 EeV, the work at Mt Chacaltaya, carried out by a collaboration of physicists from Bolivia, Japan and the USA, made use of a 60 m$^2$ shielded detector to search for showers deficient in muons, the signature of photon primaries. None was observed which we now know to be reasonable as the photon flux at the energies in question are heavily attenuated by the CMB, which had not been observed when this activity was initiated, and also the photon flux is much smaller than had been anticipated.

The second project, the initiative of J Linsley, was the construction of an array of 8 km$^2$ at Volcano Ranch, New Mexico (1770 m, 834 g cm$^{-2}$), a desert site near Albuquerque. This was the first device to cover an area of more than 1 km$^2$. In its first manifestation, the array comprised 19 plastic scintillation counters of 3.26 m$^2$ laid out on a triangular grid[6] with a spacing of 442 m. This array operated during 1959-1960. For an enlarged version, which ran from 1960 to 1963, the same detectors were spread out 884 m apart. A 20$^{th}$ detector was placed at various locations, sometimes unshielded, and sometimes covered with 10 cm of lead for muon studies.

Linsley was the first person in the shower community to face the many challenges of building such a large instrument. While for the first phase he had the full-time assistance of L Scarsi (who was visiting Rossi's group from Italy for three years on a Fulbright Fellowship), later, for the expansion, and for operation of the larger configuration, Linsley worked virtually single-handed, with only limited support to help with some of the heavier work and with some of the data-reduction. The analogue signals from the detectors were recorded by photographing oscilloscope traces. The recording equipment contained ~500 thermionic valves and was operated in the trying conditions associated with a desert climate, including dealing with rattlesnakes. Operations had to be suspended for extensive periods during the summer because of the frequent lightning storms: strikes to ground, picked-up on the earthing of the cables, damaged the electronics. The single-handed operation of the larger array was a remarkable achievement with an exposure of 25.5 km$^2$ sr year being achieved.

---

[6] There is currently discussion about the best choice of grid geometry. It is unclear why the triangular layout, which has certain advantages (Q Luce et al., Proc. 27$^{th}$ European Symposium on Cosmic Rays), was chosen for so many arrays. In a letter to Rossi written in 1986, as part of a discussion about the evolution of the arrays stemming from the MIT group, Linsley comments "On the question of rings vs. squares or some other form of lattice, after all, concentric rings make a bullseye, the classical symbol of a target". Linsley liked targets and visual allusions.

Linsley was the first to report the observation of a flattening in the number spectrum of showers at a shower size of ~$10^9$, now known as *'the ankle'*. This was interpreted as corresponding to a similar property of the energy spectrum at ~1 EeV [23], a feature which Linsley suggested might mark a cross-over between galactic and extragalactic cosmic rays. In [23] the arrival distribution of 1034 events with an average energy of over 1 EeV was presented, with the limit to any anisotropy set at <10%. The estimates of shower sizes and energies were subsequently revised following a more accurate measurement of the lateral distribution function [24] with a denser configuration of detectors laid out over 1 km$^2$. The iconic event of 100 EeV [25] was assigned a revised energy of 139 EeV [26].

In addition to developing his own group, Rossi invited scientists from Australia, Bolivia, China, France, Italy, India, Japan and elsewhere for extended stays[7]. The visits of M. Oda were of particular importance for the development of cosmic ray research in Japan.

Greisen took advantage of the scintillator factory developed at MIT, to make extensive studies over a range that extended to both lower and higher energies than achieved at the Agassiz array [27]. He used essentially the same analysis techniques as those pioneered by the MIT group. Studies of the momentum spectrum of muons were also carried out at the Cornell array.

### 3.3 Impact of work at the Geiger counter array at Harwell

The work at MIT led, indirectly, to another important development in instrumentation. At Harwell, where the UK carried out post-war atomic work, an array of Geiger counters, covering 0.61 km$^2$, was constructed *'outside the Wire'*. One of the initiators of this project was Pontecorvo who had moved to the UK following his time with Auger at Chalk River. Pontecorvo was aware of the MIT work with liquid scintillators, but, as para-terphenyl was very expensive (even now, it is more costly than printer ink), he asked J V Jelley to investigate the efficiency of the light output as a function of the quantity used [28]. Jelley found that even with no additive he could detect light – it was Cherenkov radiation – from benzene and from other liquids, including distilled water, as a muon passed through [29]. This discovery led N A Porter to develop a large water-Cherenkov detector [30] subsequently adopted for shower studies at Haverah Park and at the Pierre Auger Observatory. Porter's detector is compared (figure 1) with one of those used at the latter site. Little has changed and Pontecorvo might perhaps be regarded as the father of the water-Cherenkov detector.

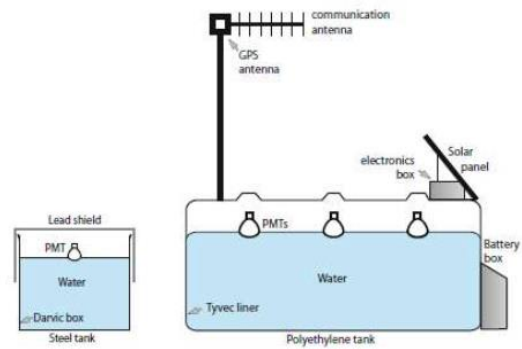

**Fig. 1.** Scale comparison of Porter's water-Cherenkov detectors with a detector at the Auger Observatory [from 31].

The array at Harwell was also used to explore the possibility of detecting Cherenkov light produced as showers traverse the atmosphere. This arose directly from a study by Blackett [32] of the essential features of Cherenkov radiation in air. He calculated that the energy thresholds for electrons and muons to produce Cherenkov radiation were 20 MeV and 4 GeV respectively, and showed that the Cherenkov radiation from cosmic rays was ~$10^{-4}$ of the night-sky background. According to Lovell [33], Blackett was at this time interested in the quantum efficiency of the eye and had concluded that extensive air showers would produce a flash of light that he should be able to see by lying down and looking upwards under suitable dark-sky conditions, an effort which Blackett undertook himself. There is no record of the outcome of Blackett's efforts, but the idea inspired Galbraith and Jelley to a successful detection of light from air showers [34]. To establish that the light detected was indeed Cherenkov radiation, further work was carried out at Pic du Midi.

The detection of air-Cherenkov radiation became a common activity at several shower arrays, notably in the Soviet Union where the method blossomed under the leadership of A E Chudakov and G T Zatsepin [35]. Important observations made there with this technique led to model-independent estimates of the primary energy of cosmic rays that produced showers of size $10^5$ to $10^6$, by Greisen [36] and S I Nikolsky [37].

---
[7] Rossi's initial efforts to have Blackett visit were thwarted, as during the McCarthy era it proved impossible for him to get a visa. Blackett did visit later when Eisenhower was President.

### 3.4 Shower studies in Japan

Cosmic-research was initiated in Japan in the 1930s by Y Nishina in his laboratory at RIKEN and flourished particularly in the period after the end of World War II when the study of nuclear physics was forbidden, an edict reinforced by the destruction of the cyclotrons at RIKEN, Kyoto and Osaka. S Tomanaga led studies of extensive air showers at Mt Norikura (2770 m) and later played a key role in establishing the Institute of Nuclear Studies (INS) in Tokyo. He was also influential in encouraging J Nishimura and K Kamata to develop three-dimensional calculations of electromagnetic cascades, work which they began after reading the Rossi and Greisen article in Reviews of Modern Physics [38], a copy of which had reached Japan just before the attack on Pearl Harbour [39].

Experimental work on showers at sea-level at INS was activated by Oda on his return from MIT in 1956. He led the construction of a series of ever-more complex arrays, culminating in one that contained 14 1 m$^2$ scintillators for measuring signal sizes and a further 5 smaller scintillators for direction measurement [40]. In addition, 8 m$^2$ of scintillator was deployed in a tunnel, 15 m deep, to measure the muon signal above 4.5 GeV. The most important insights came from combining the data from the 14 unshielded scintillators and the muon detectors. The INS group was the first to point out the information that could be derived from a study of plots of muon versus electron number, $N_\mu$ vs. $N_e$ plots in today's notation. Some data are shown in figure 2.

Large fluctuations in muon number are seen at fixed electron size, with the sharp upper boundary suggesting that the first interaction dominates the fluctuations found in showers. This type of diagram, made with greatly enhanced statistics, when combined with the detailed shower simulations that became available from the late 1960s, proved to be an important tool for extracting information on primary mass.

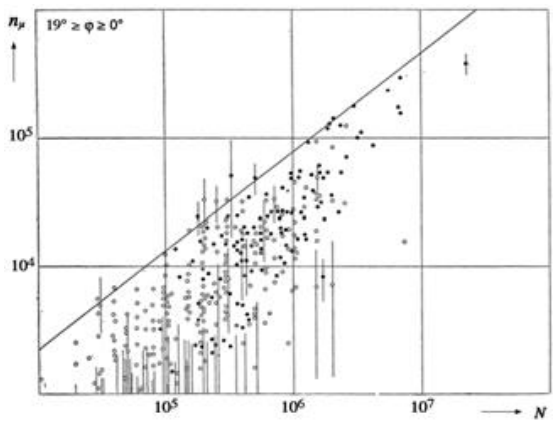

**Fig. 2.** Reconstructed muon number $N_\mu$, vs. shower size N for vertical showers [40].

These simulations also led to the recognition that, at a fixed energy, the fluctuations in electron number were much greater than those in the muon signal, so that muon number was an effective surrogate for primary energy, although of course the energy deduced was dependent upon the hadronic model adopted. Plots such as those of figure 2 led to the idea of identifying photon-initiated showers through a paucity of muons and to the efforts at Chacaltaya.

### 3.5 The Haverah Park Array

The array at Harwell was decommissioned in 1958 as the land was required for the Culham Nuclear Fusion Laboratory. A decision was taken to continue shower studies in the UK, but in the universities, and a new array was constructed on a moorland site at Haverah Park, about 25 km from Leeds. This initiative, strongly supported by Blackett, was led by Wilson who had worked with Blackett at Birkbeck in the late 1930s and moved with him to Manchester in 1937. Wilson had also interacted with both Rossi and Auger during their brief spells in Manchester [13].

Scientists from Durham University, Imperial College London and Leeds University formed the initial collaboration. Following extensive studies by the Imperial College group, led by H R Allan, the decision was taken to use water-Cherenkov detectors as the basic detecting elements. The construction and operation of the 12 km$^2$ array was the responsibility of the Leeds group. The Durham team built a magnetic spectrograph to study the momenta of muons, with the Imperial College group (later joined by researchers from Nottingham University) developing shielded detectors to measure muon signals.

In the 1970s, prompted by Greisen, the Durham group began measurements of the air-Cherenkov signal using 5 inch photomultipliers pointing upwards. The potential of using Cherenkov radiation to cross-calibrate estimates of shower energies at different arrays was explored by the Durham team, but attempts to do this for energy estimates from the Volcano Ranch and Haverah Park arrays floundered for a variety of reasons.

The problem of obtaining the energy of showers detected using an array of water-Cherenkov detectors was challenging, as it was impossible to infer the number of electrons in the events from the measurements. When the project was started, determining the shower size was still commonly seen as the first step to estimating the energy. However, Hillas [41, 42] introduced the concept of classifying the size of showers through the use of the signal at an appropriate distance from the shower axis (now called $S(r_{opt})$ and widely adopted across the community). Also, through a series of simulations [42], Hillas and colleagues established a conversion between this value and the primary energy, which was used to determine the energy spectrum [43].

Other important results obtained at the array included identifying experimentally the magnitude of the photon, electron and muon components [44], and the first detection of shower-to-shower fluctuations in showers of energy ~1EeV, deduced from a study of the risetimes of the signals from the large area water-Cherenkov detectors. This work also gave hints of a change in the Elongation Rate of showers at ~5 EeV [45, 46].

Extensive studies of radio emission in showers were also carried out under the direction of Allan (see below).

### 3.6 The Yakutsk Air Shower Array

By far the most complex, and most northerly, of the early giant arrays, was that operated by the Institute of Cosmophysical Research and Aeronomy at Yakutsk, Siberia (latitude 62° N, 105 m). This array, originally proposed by D D Krasilnikov in 1959, began taking data with 13 stations in 1967 and by 1974 had been developed to cover 18 km$^2$. A detailed description of the array, which used 58 scintillation counters as the workhorse detectors, and included 192 m$^2$ of muon detectors with a threshold of 0.5 GeV, can be found in [47]. A particularly important feature was the use of 35 photomultipliers of various sizes to measure the air-Cherenkov radiation associated with the showers. The exceptionally clear Siberian skies make the data from air-Cherenkov detectors particularly valuable and allow a unique approach to the measurement of the energy spectrum and to parameters that relate to shower development. During winter, the temperature falls to -40 C, with Cherenkov light absorption in the atmosphere negligible: only molecular and aerosol scattering are considered.

The Cherenkov-light measurements can be used to find the energy lost by electrons through ionisation, essentially using the well-known track-length integral method. At 10 EeV, this accounts for about 77% of the energy of the incoming cosmic ray and is estimated from the total Cherenkov light reaching the ground by integrating over all distances using empirically-measured lateral distribution functions. These measurements also allow access to indirect information about the longitudinal development of showers. The results on the energy spectrum [48] and the mass composition [49] of cosmic rays above 0.1 EeV are in good agreement with those from other devices using different techniques.

### 3.7 The SUGAR Array

The team from the University of Sydney, who designed 'The Sydney University Giant Air Shower Recorder (SUGAR)', introduced a totally novel method of the recording of extensive air-showers with an array of ground detectors. Before this innovation, the practice was to link detectors with cables to some common point where coincident triggers between them could be made, and the signals recorded. This method becomes impractical for areas much larger than ~10 km$^2$ as it is rarely possible to have the relatively unrestricted land access enjoyed by Linsley at Volcano Ranch. In addition, the cost of cable, their susceptibility to accidental damage, lightning strikes, and the problems of generating fast signals over many kilometres, are further handicaps. The new idea was the inspiration of M M Winn and was first described in [50]. The Sydney group proposed the construction of an array of detectors that ran autonomously, with the time of a trigger above a certain number of particles recorded with respect to a timing signal transmitted across the array.

The concept was realised in the Pilliga State Forest near Narribri (250 m), New South Wales, where 47 stations were deployed over an area of ~70 km$^2$. Most of the detectors were on a grid of 1 mile (1.6 km), with 9 on a smaller spacing to enable showers of lower energy to be studied. Time and amplitude data were recorded locally on magnetic tapes and coincidences between different stations were later found off-line at the University of Sydney.

A difficulty was that the rate of triggers of a local station, above a level low enough to be useful, is very high and the rate could not be handled with technologies then available. The solution adopted was to require coincidences between pairs of detectors, 50 m apart, buried under 2 m of earth. The concept was brilliant but it was ahead of its time in terms of the technology available. Calor gas had to be used to supply the power at each station and the reel-to-reel tape recorders proved difficult to maintain in the dusty environment. The problem of handling many magnetic tapes at a single computing site also proved to be a challenge. A further problem was that the PMTs used were 7 inches in diameter and suffered from severe after-pulsing, complicating the measurement of the signals as logarithmic time-to-height converters were used to find the amplitudes. There was also a serious difficulty in estimating the energy of events as only muons were detected and therefore there was thus total reliance on shower models with no possibility to test which model was the best because of a lack of other types of detector in the array. Attempts to overcome this issue by using a fluorescence-light detector and a small number of unshielded scintillators were unsuccessful. An energy spectrum was reported [51].

The measurement of the shower directions with a precision of a few degrees was a demonstration that the timing stamp method was effective and the most valuable data from the SUGAR array were undoubtedly from the measurements of directions, the first to be made in the Southern Hemisphere at energies above 1 EeV [52]. No anisotropies were detected.

The concept of autonomous detection was tested at Haverah Park [53] in an early attempt to devise methods to make construction of an array of ~1000 km$^2$ economical, but the innovation of autonomous detectors has had its most effective realisation first at the Pierre Auger Observatory and subsequently at the Telescope Array. The innovation of Winn lives on.

Catalogues containing details of the largest showers recorded at Volcano Ranch, Haverah Park, SUGAR and Yakutsk are available [54].

**4 The Situation in the 1980s**

It is probably not widely realised how few people were working on ultra-high energy cosmic rays over forty years ago. In the mid-1970s, the number of faculty members involved with Volcano Ranch, Haverah Park, SUGAR and Yakutsk, and the efforts focussed on the Fly's Eye detector, was around 25, so that assuming that there was 1 post doc and 1 PhD student per staff member (perhaps generous), the number involved was certainly below 80. An unfortunate consequence was that publications were almost entirely in the Proceedings of the Biennial International Cosmic Ray Conferences and consequently the work attracted little attention beyond a very limited audience. The field was not highly regarded. For example, a referee of a seminal paper submitted by Linsley and Scarsi [55] to Physical Review, wrote that it was *"on a subject mainly just of technical concern to the few specialists working on air-showers, rather than being of general and fundamental significance, i.e., much of this could equally well be a technical report privately circulated"* [56]. Fortunately, the editors saw sense and this key paper was published.

Although Linsley remained active and influential in the cosmic field until his death in 2002, data-taking at Volcano Ranch with the 8 km$^2$ array was terminated in 1963, although important work was carried out there in the 1970s with a contracted array [24]). The SUGAR effort ended in 1979. The Haverah Park work with the 12 km$^2$ array ceased in 1987, and only Yakutsk continued to operate, but with a contracted area.

Efforts by the Haverah Park group to develop an array of 100 km$^2$, seen as a pilot for an area of 1000 km$^2$, were not supported, in part because the timing signal, planned for the autonomous system was derived from TV transmissions, which were then available for only 16 hours per day. Also, the promise of the Utah Group [57] to build a fluorescence detector capable of recording $10^6$ events per year above 10 PeV, and $10^4$ per year beyond 1 EeV, imposed a planning blight on some ambitions. It is interesting to note that fewer than $10^4$ events above 1 EeV have so far resulted from the totality of efforts in Utah.

In Japan, the construction of a complex and well-provisioned array of ~1 km$^2$ started in 1983, but with the upper energy targeted at 1 EeV. Construction of the 100 km$^2$ AGASA instrument got underway in 1989, with the full array operational in 1991, the year that the first discussions of what became the Auger Observatory started.

One of the strong drivers for building the Haverah Park array was the expectation that anisotropies associated with the Galaxy would be observed. Nothing significant was found for, as we now know, the instruments of that era were over two orders of magnitude too small to observe a significant signal. What undoubtedly sustained the field was the search for the steepening predicted by Greisen and by Zatsepin and Kuzmin (the GZK effect). Searching for this feature was seen as a significant target and shows the importance of having theoretical predictions when one is competing for money. Without the GZK goal, the possibility of gaining continuing support for any large projects to study air showers would have been small.

The absence of other predictions made it difficult for J W Cronin and me to argue for construction of an array of the size of the Auger Observatory. The hardest question that we had to face, and we were asked it many times, was *'why do you want to make the array so large?'* However, had we not built 3000 km$^2$ we would still be waiting to observe significant anisotropies. Phenomenologists must recognise that they have a key role to play in predicting what might be seen, and therefore of the scale of the instruments that are needed.

**5 Work in Europe in the 1990s**

Although, in the 1990s, activities in the field of ultra-high cosmic rays slowed internationally, work at lower energies blossomed in Europe, in both Germany and Italy. Additionally, a claim from the Kiel group in 1983 [58] of the observation of PeV gamma rays from Cygnus X-3, although now regarded as likely to have been incorrect, proved to have enormous impact on the development of both the TeV γ-ray and UHECR scenes.

In Italy, a group under the leadership of G Navarra constructed and operated an array, EAS-TOP, at Campo Imperatore (2005 m) as part of the National Gran Sasso Laboratory. The facility operated from 1989 to 2002 and, in its final configuration, consisted of 35 10 m$^2$ scintillation counters spread over ~$10^5$ m$^2$, together with a 144 m$^2$ calorimeter, designed for measurements of muons and hadrons, with the target of getting information on primary mass in the region of the knee [59]. Of particular importance are the data on the anisotropy of cosmic rays in the 0.1 to 1.0 PeV range [60]. In addition to a highly-significant detection of the long-sought Compton-Getting effect, the anisotropic feature that is expected due to the motion of the Earth around the Sun, a first harmonic amplitude in sidereal time, based on an analysis of 2.1 x $10^9$ events at ~0.1 PeV, was observed with a significance of over 10 sigma. The amplitude of the sidereal signal was 2.6 x $10^{-4}$ at a phase of ~6°. At higher energies (around 0.4 PeV) a morphology change in the anisotropy was observed for the first time, with the sidereal signal having a larger amplitude ($6.4 \times 10^{-4}$) and a different phase (~200°).

Contemporaneous with the work at Campo Imperatore, a team led by G Schatz, constructed and operated an array at the Institute für Kernphysik in Karlsruhe (110 m) in Germany. The instrumentation was spread over 200 x 200 m$^2$, with 250 shielded and unshielded detectors, a muon tracking device and a large calorimeter [61]. The main aim of this project was to determine features of the cosmic ray spectrum, and the variation of the primary mass, across the knee region. Detailed results were reported, with the knee position and the mass composition extracted being dependent upon the hadronic physics model adopted. A major result from the group was the demonstration, *without recourse to models*, that the number spectrum of muons steepened around the knee, but only in showers that were rich in electrons (i.e. showers produced by light nuclei) [62]. This work finally established that the knee feature in the energy spectrum, first demonstrated by Kulikov and Khristiansen [18] in 1959, was due to astrophysics and not to a change in hadronic interactions, thus putting to rest a debate that had lasted for over 40 years. I find such *model-independent conclusions* very compelling: trends are demonstrated that can be interpreted readily and unambiguously with an elementary understanding of the principles of shower development.[8] Another important product from the work in Karlsruhe was the creation of the CORSIKA code for shower simulations [63], now widely used across the cosmic ray community. The report has been cited over 1000 times.

Towards the end of the periods of running of the EAS-TOP and KASCADE arrays, the groups merged to build a larger device, KASCADE-Grande, at the KASCADE site [64]. This array, spread over 700 x 700 m$^2$, contained an additional 37 10 m$^2$ scintillation detectors. A major result was the discovery, *again without resorting to assumptions about hadronic models*, of a second knee at around 0.1 EeV [65]. This spectral break is suggestive of iron nuclei having attained a rigidity such that they can no longer be contained within the Galaxy.

It is worth noting that around 60 people signed papers from the KASCADE-Grande effort, not so different from the total number involved in UHECR activities in the 1970s. With this project, the era of the large collaborations needed for effective research in the field had arrived.

**5.1 The impact of Cygnus X-3**

From the mid-1960s, there was considerable interest in studying the central regions of air showers. This was possible, of course, only at relatively low energies (~1 PeV), but the observation of showers with single and multiple cores was pushed particularly strongly by the Sydney group [66] who used a chessboard configuration of 64 scintillators at the centre of an array of 300 Geiger counters. In addition to a claim for a flux of deuterium above 1 PeV, they presented evidence for anomalously large transverse momenta of pions in the showers. Similar effects were reported from the INS array were a neon hodoscope system was used to investigate the core region [67]. It was eventually recognised that many of the sub-cores seen in the detectors were the products of showering of particles from denser regions of the roofs covering the detectors. At this time, the power of shower simulations to investigate the impact of such configurations lay some way off.

In Kiel, West Germany, a group under the leadership of J Trümper made a particularly careful study of cores of showers. A neon hodoscope of 34 m$^2$, under a very light roof, was used along with an array of 16 1 m$^2$ scintillation counters. No anomalous effects were detected and the claims of large transverse momenta, based on core structure, slowly evaporated over the next decade. After Trümper moved to Tübingen and a career in X-ray astronomy, operation of the Kiel array continued, and in 1983 Samorski and Stamm [58] reported evidence of a signal, attributed to γ-rays of ~1 PeV from Cygnus X-3. In addition to a 4.4 sigma excess in the region around the object, a variation with the 4.8$^h$ modulation characteristic of the source was noted. Similar time variations had previously been seen at lower energies with TeV telescopes, and subsequently there was an apparent confirmation of the effect by the Haverah Park group at ~ 1 PeV [68].

The evidence never strengthened but the excitement generated by the claim led a number of eminent particle physics to join the field. While several eventually moved to TeV astronomy, the most distinguished of the defectors remained in cosmic-ray physics. This was Cronin, who led the construction of the CASA array. Located at the Fly's Eye site in Utah, it covered 0.23 km$^2$ with 1089 scintillation detectors [69]. The array, which was built with the goal of detecting γ-rays from Cygnus X-3 several times each day, also contained a muon detector of 2560 m$^2$, which was used as a veto against charged cosmic rays. Although no γ-rays were ever detected, Cronin's appetite for working in cosmic rays was whetted and, in my view, without his singular input and dedication, it is unlikely that the Auger Observatory would ever have been developed. Life would have been very different for many of us without Cygnus X-3.

# 6 Subsequent developments

## 6.1 Hardware

It may be surprising, and perhaps even a little depressing, to realise that the major detection techniques, the scintillation counter and the water-Cherenkov detectors, first used for cosmic ray work in the 1950s, still retain such a central role in our work. Of course, there have been developments: in particular, photomultipliers and scintillator material are hugely cheaper than in the early days, optical fibres now aid

---

[8] An identical method had been introduced some years earlier by the CASA-MIA Collaboration to reach the same conclusion, although from a smaller number of events (M A K Glasmacher et al. Astroparticle Physics 12 1 (1999)).

light collection from large areas, and photomultipliers have much higher photocathode efficiencies.

The development of autonomous detector arrays, following the innovation of the Sydney group in the 1960s, has been of immense importance. Fluorescence detection (discussed in the talk by P Sokolsky) and measurements of radio emission have come to the fore, but it is worth recalling that the use of these were first discussed in the 1960s. In the case of the fluorescence method, it took until the late 1980s before a measurement of the energy spectrum was reported from it, and successful exploitation of the radio technique took even longer.

In 1965, at the ICRC in London, Greisen gave an invited paper, still worth reading, that was insightful and influential [70]. Amongst the topics he discussed was the detection of air showers using the radio method, which had just been demonstrated [71]. Greisen commented: *'The technique is barely in its infancy.... We feel confident it is a significant breakthrough and further study will reveal ways of obtaining information about showers not available by other means'*. His prediction would eventually turn out be correct but it was to be an infancy lasting nearly 40 years as the method fell out of favour around 1975, largely because of the inability to monitor the geo-electric field adequately at that time. The approach was revived early in the 21$^{st}$ century and is now a key tool, being exploited at LOFAR to study the depth of shower maximum and at the Auger Observatory for the same purpose, but also to improve estimates of the primary energy. The technique is also being applied in several projects targeted at the detection of very high-energy neutrinos.

Despite several efforts over the years, the radar echo method has yet to be successful.

### 6.2 Developments from growth in computing power

It is hard to imagine how our field could have advanced without the availability of the Monte Carlo technique, but it took around 20 years from the invention of the method by Ulam and von Neumann for sufficient computer power to be available for calculations relevant to air showers to become practical. Results from the first such work, using computers, were presented at the London ICRC in 1965. Those reported by Hillas were carried out at Leeds University on a KDF9 mainframe machine with 64 k of memory weighing about 5 tonnes, housed in a disused church and costing over €35M in today's money. Technology has advanced and the extensive use of the CORSIKA code had already been mentioned.

Adoption of the ideas of Artificial Intelligence (AI) and Neural Networks, now gaining wide use across our field, have also had a long gestation with early claims for the power of these concepts often over-hyped. Historians speak of two 'AI winters', 1974-1980 and 1987-1993, during which funding for AI research largely dried up. In 1991, for example, massive Japanese investment was curtailed when objectives, such as using computers to have a casual conversation, had not been met, a target only achieved in 2010. During 1991, around 300 commercial AI companies were shut down or went bankrupt.

Interestingly, during the second winter, AI was being used to address problems of importance to the astroparticle physics community. Halzen, Vazquez and Zas [72] and Reynolds [73], developed Neural Network techniques to analyse images from TeV gamma-ray telescopes. Perrett and van Stekelenborg, part of the team working in the Bartol Institute on analysis of data from the SPASE experiment at the South Pole, used the Neural Net approach to improve the speed of shower reconstruction [74].

### 6.3 Hadronic Physics

A big surprise, at least to me, has been that it has proved possible to obtain interesting results relating to hadronic physics at energies well-beyond those reached at the LHC from air-shower studies. Data from the Auger Observatory have been used to measure the p-air cross-section at a centre-of-mass energy of 57 TeV [75]. An excess in the number of muons, over that expected from hadronic models, which describe LHC data has also been uncovered [76].

Such work was not foreseen when the project was planned. However, if it does prove possible to identify the mass of individual events, as is envisaged with the extension of the Auger Observatory, this line of research might well develop strongly.

## 7 The Pierre Auger Observatory and the Telescope Array

The idea of building a truly giant shower array was developed during the 1990s and evolved into the Pierre Auger Observatory [77]. The goal of the project was to measure the properties of the highest energy cosmic rays with unprecedented precision. Data taking began on 1 January 2004 with a partial deployment of the instrumentation. The full Observatory of 1600 12 m$^2$ water-Cherenkov detectors on a 1500 m triangular grid spread over 3000 km$^2$ and overlooked by four fluorescence detectors, was completed in June 2008. Subsequently, 30 m$^2$ of muon detector was added, along with a number of radio antenna. Phase I of the effort ended on 31 December 2020. For phase II, named AugerPrime, 4 m$^2$ scintillator are being placed above each water-Cherenkov detector, with radio antennas added at every station. A major target of phase II is the determination of the mass of the primary particles individually rather than on an average basis.

The only giant array operating during the 1990s was the AGASA instrument, which covered ~100 km$^2$. Striking results were reported using data from it including a claim that there were events with energies well-beyond the steepening expected from the GZK effect [78] and a report of clustering of events above 10 EeV [79]. To check these results with higher statistics, the decision was taken to build the Telescope Array, a hybrid instrument with 507 scintillators spread across 700 km$^2$ on a 1200 m square grid [80], and overlooked by three fluorescence detectors [81]. Data-taking at the Telescope Array started in 2007.

These two observatories are on-going projects and, in a paper devoted to history, no further discussion is offered.

**7 Concluding remarks**

Study of cosmic rays with air-showers across the energy spectrum from 0.1 PeV to the very highest energies is in a lively state. What I have tried to show is the evolution that has led to the present vibrancy of our field. Our present position owes much to the input of a small number of iconic figures, most notably, in my view, Auger, Blackett, Chudakov, Greisen, Hillas, Jelley, Linsley, Rossi and Zatsepin. We continue to exploit the ideas and techniques which they developed but with much better financial support. We should never forget that we are standing on the shoulders of giants.

I am very grateful to the organisers of UHECR2022 for inviting me to indulge in one of my fascinations – the history of our field. Parts of the presentation above draw on the paper by Karl-Heinz Kampert and myself [31]. I recall how enjoyable it was to work with him on that project. I would also like to thank Valerie Higgins (archivist at FNAL) for her help during a visit to the Linsley archives in July 2019.